\documentclass[fleqn,usenatbib, letters, onecolumn]{mnras}
\usepackage{mathptmx}
\usepackage[T1]{fontenc}
\usepackage{ae,aecompl}
\usepackage{rotating}

\usepackage{graphicx}	% Including figure files
\usepackage{amsmath}	% Advanced maths commands
\usepackage{amssymb}	% Extra maths symbols
\usepackage{epstopdf}

\title[Switches between accretion structures during flares]{Switches between accretion structures during flares in 4U 1901+03}

\author[Long, Ji et al.]{%
	L. Ji$^{1}$ \thanks{E-mail: ji.long@astro.uni-tuebingen.de},
	L. Ducci$^{1, 2}$,
	A. Santangelo$^{1, 3}$,
	S. Zhang$^{3}$,
    V. Suleimanov$^{1, 4, 6}$,
    \newauthor
	S. Tsygankov$^{5, 6}$,
	V. Doroshenko$^{1, 6}$,
	A. Nabizadeh $^{5}$,
	S. N. Zhang$^{3,7}$,
	M. Y. Ge$^{3}$
% ...
\\
$^{1}$ Institut f\"ur Astronomie und Astrophysik, Kepler Center for Astro and Particle Physics, Eberhard Karls Universit\"at, Sand 1,\\ 72076 T\"ubingen, Germany\\
$^{2}$ ISDC Data Center for Astrophysics, Universit\'e de Gen\`eve, 16 chemin d'\'Ecogia, 1290 Versoix, Switzerland\\
$^{3}$ Key Laboratory for Particle Astrophysics, Institute of High Energy Physics, Beijing 100049, China\\
$^{4}$ Kazan (Volga region) Federal University, Kremlevskaya str. 18, 42008 Kazan, Russia,\\
$^{5}$ Department of Physics and Astronomy, University of Turku, FI-20014 Turku, Finland\\
$^{6}$ Space Research Institute of the Russian Academy of Sciences, Profsoyuznaya Str. 84/32, Moscow 117997, Russia\\
$^{7}$ University of Chinese Academy of Sciences, Chinese Academy of Sciences, Beijing 100049, People’s Republic of China\\
$^{8}$ Department of Astronomy, Beijing Normal University, Beijing 100088, People’s Republic of China\\
$^{9}$ Department of Astronomy, Tsinghua University, Beijing 100084, China\\
$^{10}$ Department of Physics, Tsinghua University, Beijing 100084, China\\
$^{11}$ Department of Engineering Physics, Tsinghua University, Beijing 100084, China\\
$^{12}$ Key Laboratory of Space Astronomy and Technology, National Astronomical Observatories, Chinese Academy of Sciences, \\
Beijing 100012, China\\
$^{13}$ College of physics Sciences \& Technology, Hebei University, Baoding 071002, Hebei Province, China\\
}

\date{Accepted XXX. Received YYY; in original form ZZZ}

\pubyear{2019}

\begin{document}
\label{firstpage}
\pagerange{\pageref{firstpage}--\pageref{lastpage}}
\maketitle

\begin{abstract}
We report on our analysis of the 2019 outburst of the X-ray accreting pulsar 4U 1901+03 observed with \textit{Insight-HXMT} and \textit{NICER}.
Both spectra and pulse profiles evolve significantly in the decaying phase of the outburst. 
Dozens of flares are observed throughout the outburst. They are more frequent and brighter at the outburst peak.
We find that the flares, which have a duration from tens to hundreds of seconds, are generally brighter than the persistent emission by a factor of $\sim$ 1.5.
The pulse profile shape during the flares can be significantly different than that of the persistent emission.
In particular, a phase shift is clearly observed in many cases.
We interpret these findings as direct evidence of changes of the pulsed beam pattern, due to transitions between the sub- and super-critical accretion regimes on a short time scale.
We also observe that at comparable luminosities the flares' pulse profiles are rather similar to those of the persistent emission. 
This indicates that the accretion on the polar cap of the neutron star is mainly determined by the luminosity, i.e., the mass accretion rate.

\end{abstract}


%%%%%%%%%%%%%%%%%%%%%%%%%%%%%%%%%%%%%%%%%%%%%%%%%%

%%%%%%%%%%%%%%%%% BODY OF PAPER %%%%%%%%%%%%%%%%%%

\section{Introduction}
Accretion-powered pulsars are highly magnetised neutron stars with a magnetic field $B$ of $\sim 10^{12}$\,G, and a stellar companion in binary systems.
The accreted matter is channelled by the magnetic field onto the magnetic poles of the neutron star, resulting in coherent pulsations, if spin and magnetic axes are misaligned.
The pulse profile shape is thought to depend on luminosity, since the changes of the accretion structures and radiation transfer close to the polar cap depend on the accretion rate.
The variation of pulse profiles therefore reflects the changes of the radiation beam patterns \citep{Basko1976, Becker2012, Mushtukov2015}.
As widely discussed in the literature, the observation and interpretation of the evolution of pulse profiles with luminosity and energy play an important role in understanding accretion physics on magnetised objects. In fact, a large number of papers has been published in this field by using extensive data observed in many outbursts of pulsars at different luminosity states \citep[see, e.g. ,][]{Parmar1989,Mukerjee2000, Sasaki2012, Reig2013, Wilson2018,Doroshenko2019, Ji2019}.
In addition, some pulsars show luminosity-dependent short-term variabilities that resemble the  long-term behaviour along outbursts.
For instance, \citet{Klochkov2011} and \citet{Vybornov2018} reported the evolution of the spectral shape and the cyclotron line energy with luminosity by using pulse-amplitude-resolved analysis.
Dramatic increases of the flux on a short time scale, referred here as "flares", are observed in the lightcurves of some accreting pulsars such as A0535+26, GRO J1744-28, SMC X-1, Swift J1626.6-5156, LMC X-4 and A0538-66 \citep{Caballero2008, Woods2000, Moon2003, Reig2008, Doroshenko2012, Shtykovsky2018, Ducci2019}.
They are generally thought to be produced by instabilities (e.g., Rayleigh-Taylor instability and magnetospheric instability) in the accretion disc or clumpy stellar winds from high mass donor companions \citep{Taam1988, Apparao1991, Cannizzo1996, Cannizzo1997, Moon2003, Postnov2008}.
We note that the flares in these sources are very different from each other, and are also significantly distinct from the flares of 4U 1901+03 reported below.

4U 1901+03 is a transient X-ray pulsar discovered with \textit{Uhuru} and \textit{Vela\ 5B} during its outburst in 1970 \citep{Forman1976,Priedhorsky1984}.
In 2003 its second outburst was observed by the \textit{Rossi\ X-ray\ Timing\ Explorer (RXTE)}. Coherent pulsations were found with a period of $\sim$2.76\,s \citep{Galloway2003a}.
The distance to the source is $3.0^{+2.0}_{-1.1}$\,kpc measured by \textit{Gaia} \footnote{source\_id = 4268774695647764352} \citep{Bailer2018}.
The spectral shape of the source has been generally described as a power-law model with a high energy cutoff (cutoffpl).
In addition, a prominent 10\,keV absorption feature was used to model the spectrum during the outburst.
 \citet{Reig2016,Mereminskiy2019} suggested that the feature might be a cyclotron resonant scattering feature (CRSF), although its origin is still debated.
In the 2003 outburst, flares were identified and studied by \citet{James2011}.
These flares had a duration of $\sim$100-300\,s and showed clearly spectral changes. In addition, the flares were more frequent and brighter at the peak of the outburst.
Using the pulse-amplitude-resolved analysis, they reported that the pulse profile is intensity-dependent, and appears to be similar between flaring and non-flaring episodes.
As detected by \textit{MAXI} and \textit{Swift}/BAT, 4U 1901+03 entered a new outburst in February 2019 \citep{Nakajima2019, Kennea2019}.
In this paper, we focus on the flares detected with \textit{Hard X-ray Modulation Telescope} (\textit{Insight-HXMT}) and \textit{Neutron star Interior Composition Explorer} (\textit{NICER}).
The aim of this paper is to establish a link between the pulse profile of flares and that of non-flare emissions during the outburst evolution, which has not been reported in the literature.
This paper is arranged as it follows: We introduce data reduction in Section 2, and present our results in Section 3. We discuss the observed phenomenology in Section 4.

\section{Observations and data analysis}
\begin{figure}
\centering
	\includegraphics[width=0.6\columnwidth]{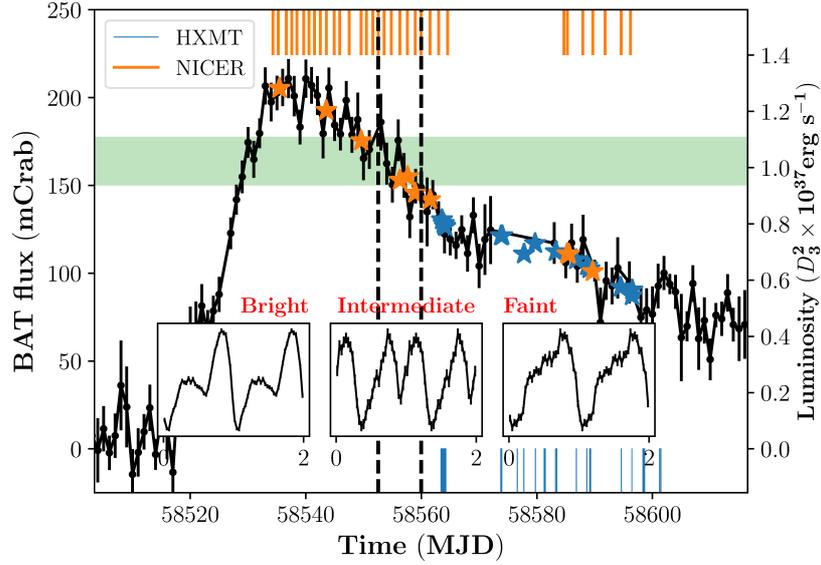}
    \caption{The lightcurve (black line) of the outburst of 4U 1901+03 in 2019 observed with \textit{Swift}/BAT. The vertical solid lines represent the \textit{NICER} and \textit{Insight-HXMT} observations in orange and blue, respectively. The stars show the time when flares are detected.
    The pulse profiles of non-flare emissions generally have three typical shapes depending on the luminosity. We show examples in insets and the corresponding time intervals separated by vertical dashed lines.
    The horizontal green strip represents the inferred flux at which the pattern 3 (see Figure~\ref{fig:sketch}) is shown in the outburst.
    }
    \label{fig:lc}
\end{figure}

The latest outburst of 4U 1901+03 was observed by \textit{Swift}/BAT and \textit{MAXI} in February 2019 \citep{Kennea2019, Nakajima2019}.
This outburst had a duration of $\sim$ 100 days, with a peak flux of $\sim$200\,mCrab at 15-50\,keV.
We show the long-term \textit{Swift}/BAT monitoring lightcurve in Figure~\ref{fig:lc}.
In this paper we mainly focus on the timing analysis of the flares (see Figure~\ref{fig:hardness_flare} for an example) observed with \textit{NICER} and \textit{Insight-HXMT}.
\textit{NICER} is an X-ray telescope onboard the International Space Station.
{In this paper, we used events in the 1-10\,keV energy range to avoid the optical loading at lower energies.
We performed data reduction using the recommended pipeline {\sc nicerl2} in {\sc heasoft} v6.25 \footnote{https://heasarc.gsfc.nasa.gov/docs/software/heasoft/}.}
\textit{Insight-HXMT} is a Chinese telescope, which consists of low energy detectors (LE; 1-10\,keV), medium energy detectors  (ME; 10-30\,keV) and high energy detectors (HE; 25-250\,keV).
{In order to have a high signal-to-noise ratio, here we only used the energy ranges of 1-10\,keV and 10-20\,keV for the LE and ME, respectively}.
We performed the analysis by following the official user guides \footnote{https://heasarc.gsfc.nasa.gov/docs/nicer/nicer\_analysis.html}
\footnote{http://www.hxmt.org/index.php/dataan/fxwd/347-the-hxmt-data-reduction-guide-v1-01}
using the software {\sc hxmtsoft} v 2.01 \footnote{http://www.hxmt.org/index.php/dataan}.
The screening criteria to generate the good time intervals of \textit{Insight-HXMT} data is that the elevation angle > 10 degree; the geomagnetic cutoff rigidity > 8 GeV; the pointing offset angle < 0.1 degree; at least 600\,s away from the South Atlantic Anomaly (SAA).
The background of \textit{Insight-HXMT} was estimated by using standalone {\sc python} scripts ({\sc lebkgmap} and {\sc mebkgmap}) and was subtracted off in the following analysis.
The time scale of the variability of the background is of the order of ks, which does not significantly influence our analysis.
There are 37 \textit{NICER} observations and 19 \textit{Insight-HXMT} observations for the source.
As shown in Figure~\ref{fig:lc}, \textit{NICER} had a higher observation cadence when the source was bright, whereas \textit{Insight-HXMT} performed more observations in the decay phase of the outburst.
This complementary monitoring allows us to extensively study flares at different luminosity states.
In both \textit{NICER} and \textit{Insight-HXMT} observations, pulsations could be significantly found ($P_{\rm s}$ $\sim$ 2.72\,s) by using the phase-connection technique \citep{Deeter1981}.
The detailed spinning-up evolution of this source and an updated ephemeris have been reported by \citet{Tuo2019} and adopted in this paper.
In the following analysis, barycentric and binary corrections have been performed when studying pulse profiles.
Here the pulse profiles were extracted by folding the events for given energy ranges, e.g., 1-10\,keV for \textit{NICER} and 10-20\,keV for \textit{Insight-HXMT} by assuming the reference time (t=0) as the point of $\rm Phase$=0.

\begin{figure}
	\centering
	\includegraphics[width=0.5\linewidth]{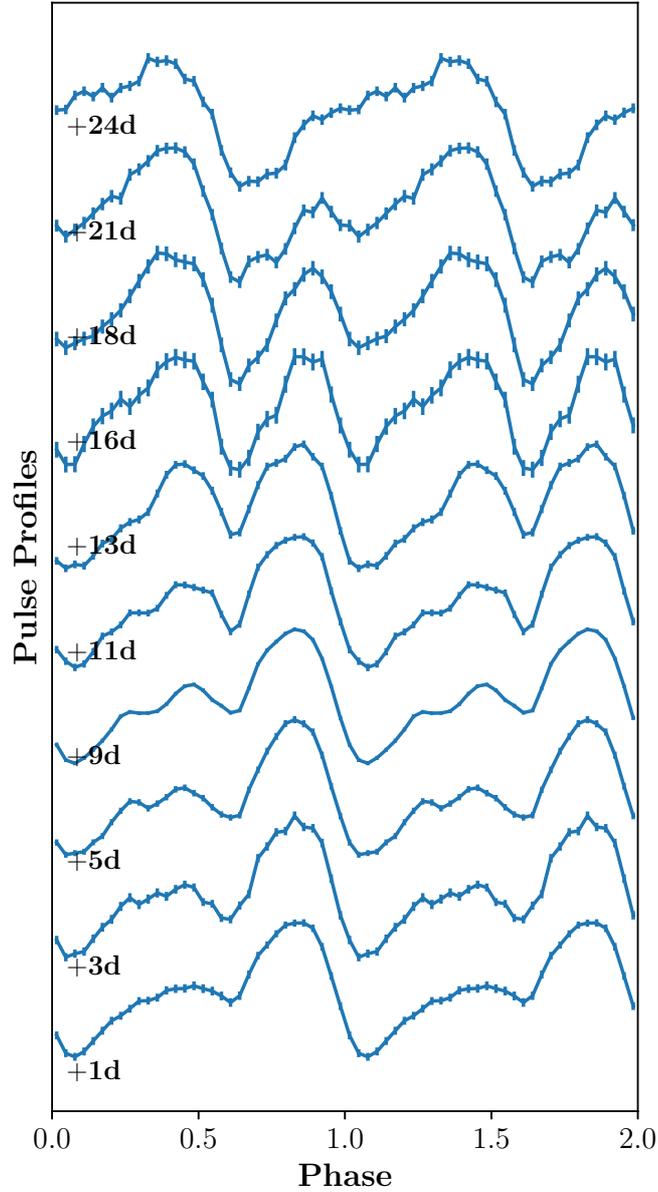}
	\caption{The evolution of pulse profiles of the non-flare emissions observed with \textit{NICER} in the energy range of 1-10\,keV after MJD 58540. For clarity, we aligned pulse profiles by using the cross-correlation technique.
    The pulse profile shape evolves evidently from bottom to top with the decreasing of the luminosity.
   }
	\label{fig:ppev}
\end{figure}

\begin{figure}
\centering
	\includegraphics[width=0.6\columnwidth]{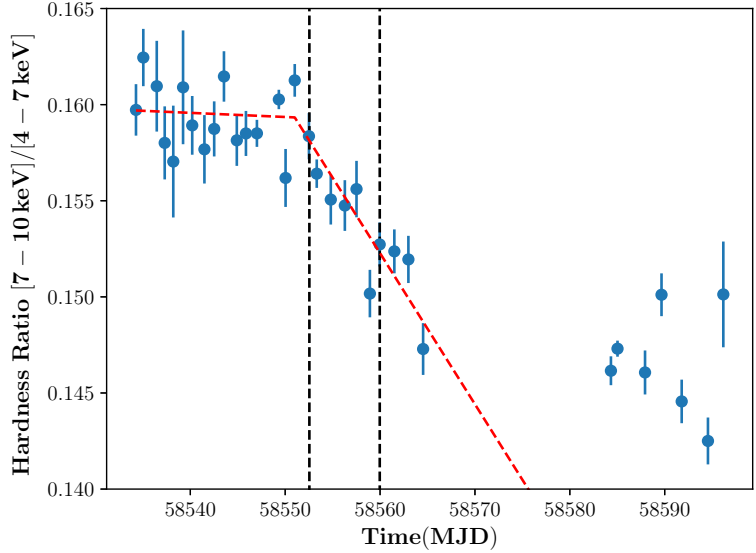}
    \caption{Hardness ratio evolution of 4U 1901+03 observed with \textit{NICER}. The red dashed line represents a fitting with a broken power-law model before MJD 58570.}
    \label{fig:hardness}
\end{figure}

\section{results}
\subsection{Hardness and pulse profile during the outburst}
First we investigated the spectral evolution of the source through the hardness changes.
Hardness here is defined as the ratio of \textit{NICER} count rates between two energy bands, namely, 4-7\,keV and 7-10\,keV.
These energy ranges are widely used in studies of accreting pulsars \citep[see, e.g.,][]{Reig2013}.
We show the result in Figure~\ref{fig:hardness}.
{We fitted the hardness evolution before MJD 58570 with a power-law function and a broken power-law function, and found that the latter was better at a confidence level of > 4$\sigma$ estimated by F-test.}
The hardness appears to be constant around the outburst peak until $\sim$ MJD\,58555, and then smoothly decreases in the decay phase of the outburst.

In addition, we have found that the pulse profile shape of the non-flare emission changes along the outburst.
We show examples in Figure~\ref{fig:lc} and \ref{fig:ppev}.
More details on pulse profiles and the corresponding pulsed fraction will be reported in a separated paper (Nabizadeh et al. in preparation).
There are generally three types of pulse profiles: when the source is bright ($\gtrsim$ $10^{37} D_{3}^2 \,{\rm ergs\ s^{-1}}$, where $D_{3}$ is the distance in the units of 3\,kpc; before MJD 58550), the pulse profile consists of a main peak and another small peak; when the source is faint  ($\lesssim$ $10^{37} D_{3}^2 \,{\rm ergs\ s^{-1}}$; after MJD 58560), the pulse profile has only one broad peak, similar to some other sources at the low luminosity state \citep[see, e.g.,][]{Devasia2011, Malacaria2015}; in the intermediate state (MJD 58550 < time < MJD 58560), the pulse profile is complex and dramatically evolves, approximately showing double peaks or one single peak interrupted by a "notch".
We note that the time interval when pulse profiles significantly change well coincides with that when hardness starts to decrease.

\subsection{Lightcurves and hardness ratios of flares}
\begin{figure}
\centering
	\includegraphics[width=0.6\columnwidth]{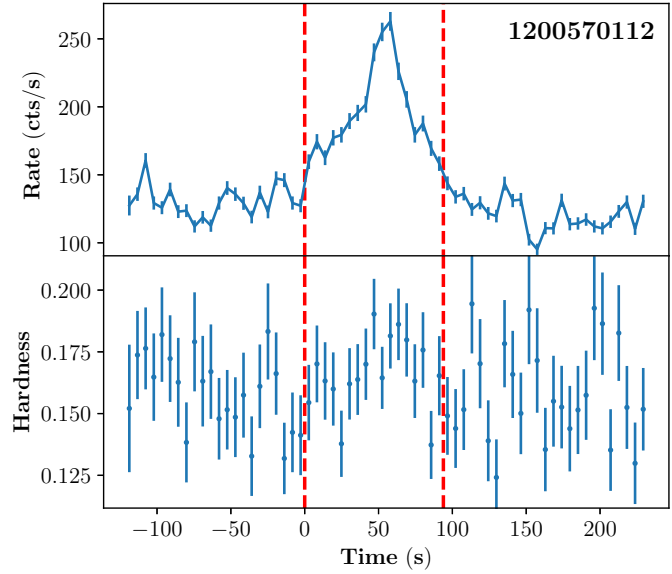}
    \caption{An example of the hardness (7-10\,keV/4-7\,keV) evolution during a flare. Vertical red lines represent the start and stop of the flare. }
    \label{fig:hardness_flare}
\end{figure}

The most prominent feature in the lightcurves of 4U 1901-03 is its aperiodic variability, showing flares, with the time scale of tens to hundreds of seconds \citep{James2011}.
Flares show a strong variability, and are on average brighter than the persistent emission by a factor of $\sim$1.5 (see an example in  Figure~\ref{fig:hardness_flare}).
We searched for the variability of count rates of \textit{NICER} (1-10\,keV) and \textit{Insight-HXMT} (LE and ME, 1-10\,keV and 10-20\,keV) lightcurves, and selected flares only if their peaks are brighter than nearby non-flare emissions at a confidence level of >3\,$\sigma$.
The confidence level is estimated by fitting the non-flare count rate distribution with a gaussian function.
Such a selection criteria leads to a significant identification of flares in our sample with a confidence level of at least 7$\sigma$, assuming a null hypothesis that flares are caused by the statistical fluctuation of non-flares emissions \citep{Ji2014}.
We defined the starting and ending points of flares as the time when the count rate is higher than the non-flare emissions at a 90\% confidence level.
In practice, to avoid flare-like background events of \textit{Insight-HXMT}, we identified a flare only if it was detected with both LE and ME detectors or a {significant change of pulse profiles} was present (see below).
The good time intervals of ME and LE could be different, and the former generally has more exposure.
Thus, for \textit{Insight-HXMT} we mainly used the ME as a reference to track the flares.
The available LE data were also used in determining the pulse shift (see  below).
We note that the terminology of "flares" is just used to refer to short-term time intervals with a higher count rate phenomenologically, which does not reflect an underlying accretion physics distinct from the non-flares.
Based on our selection criteria, most of the flares were identified around the outburst peak because of the high flux, but when the source is faint we were unable to discern them from fluctuations of the persistent emission.
A similar result was reported by \citet{James2011}.
In total, we selected 11 and 16 flares into our sample from \textit{NICER} and \textit{Insight-HXMT} observations, respectively.
The full data set is included in Appendix~\ref{appendix} and we only show some representative flares in the main text.

We then investigated the hardness evolution during the flares, which revealed no significant variation.
We note that it is mainly due to poor statistics, because the expected hardness variability during flares is $\sim$ 0.005 \footnote{if we assume that the hardness-flux relation of flares is comparable to that shown in the outburst evolution (Figure~\ref{fig:hardness}).
}, which is much smaller than the statistical error.
We show an example of a flare around the outburst peak in Figure~\ref{fig:hardness_flare}.

\subsection{Pulse profile evolution of flares}
\begin{figure*}
\centering
	\includegraphics[width=0.4\columnwidth]{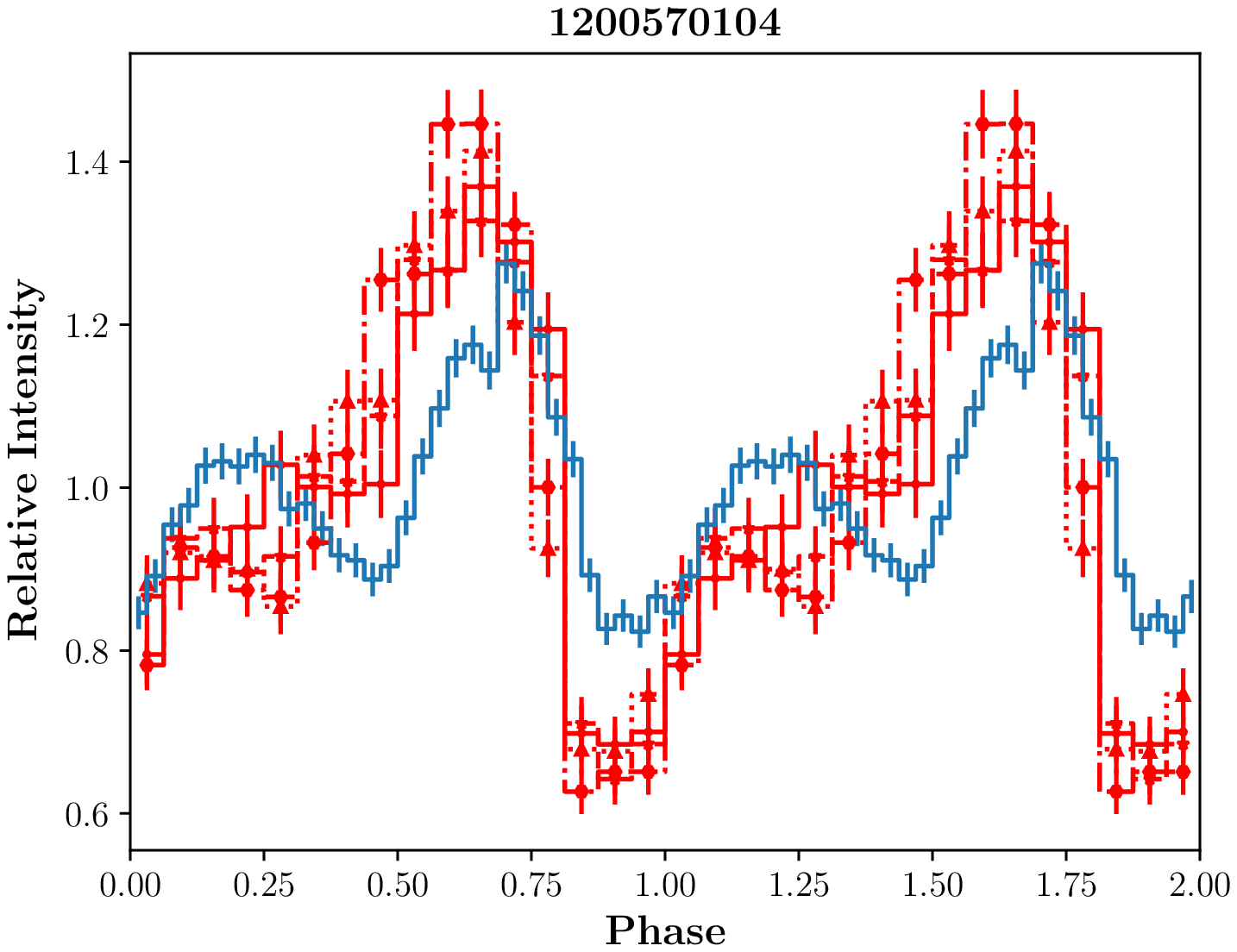}
	\includegraphics[width=0.4\columnwidth]{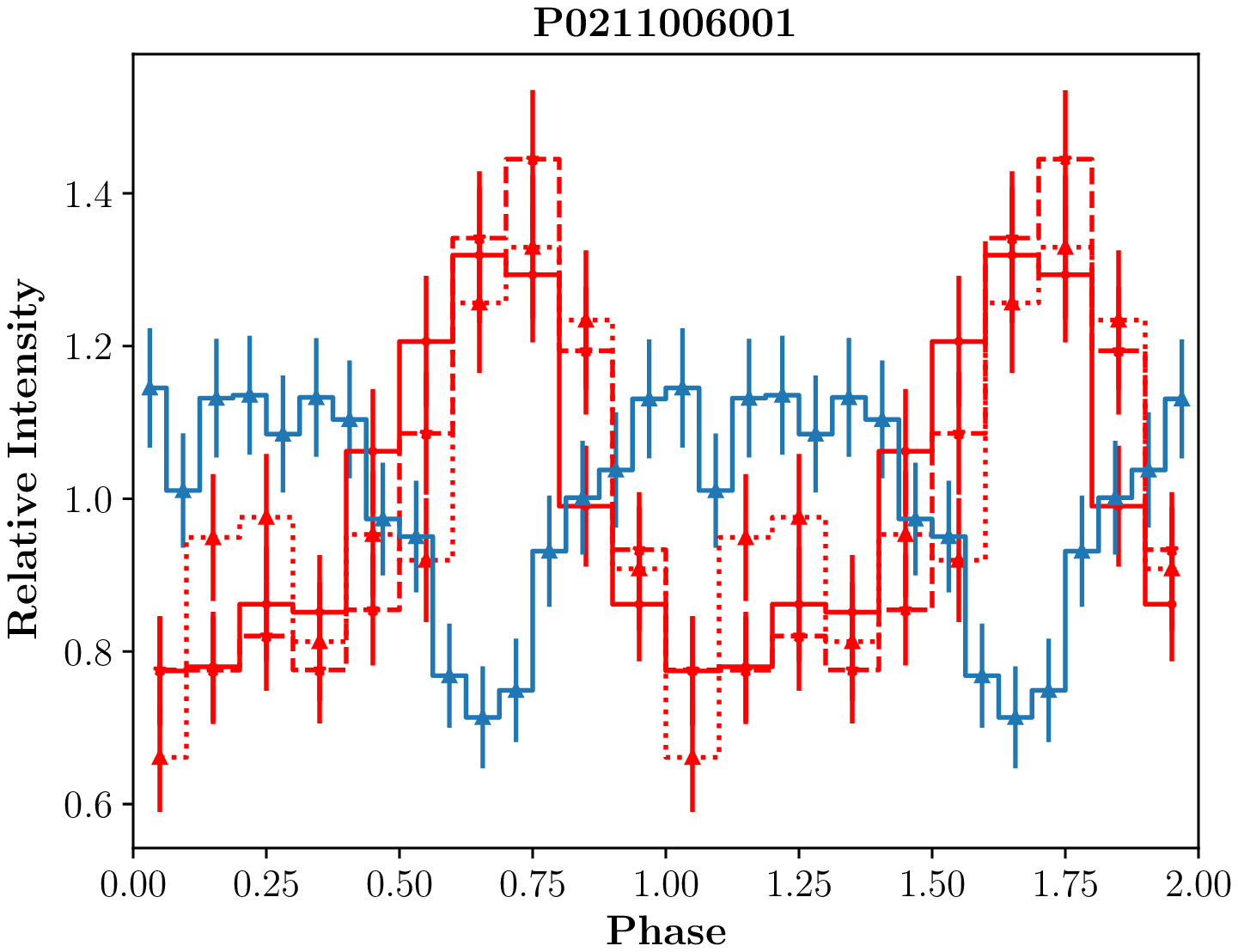}
    \caption{
    {Examples of pulse profiles observed with \textit{NICER} (1-10\,keV; left) and \textit{Insight-HXMT}/ME (10-20\,keV; right). The blue and red lines represent pulse profiles of non-flares and flares. The flares are divided into several segments and shown in different line styles, which are quite similar to each other. Therefore, for the statistical reason, we combine all events of flares as a whole to study their pulse profiles in this paper.}
    }
    \label{fig:flares_sub}
\end{figure*}

\begin{figure*}
\centering
	\includegraphics[width=0.45\columnwidth]{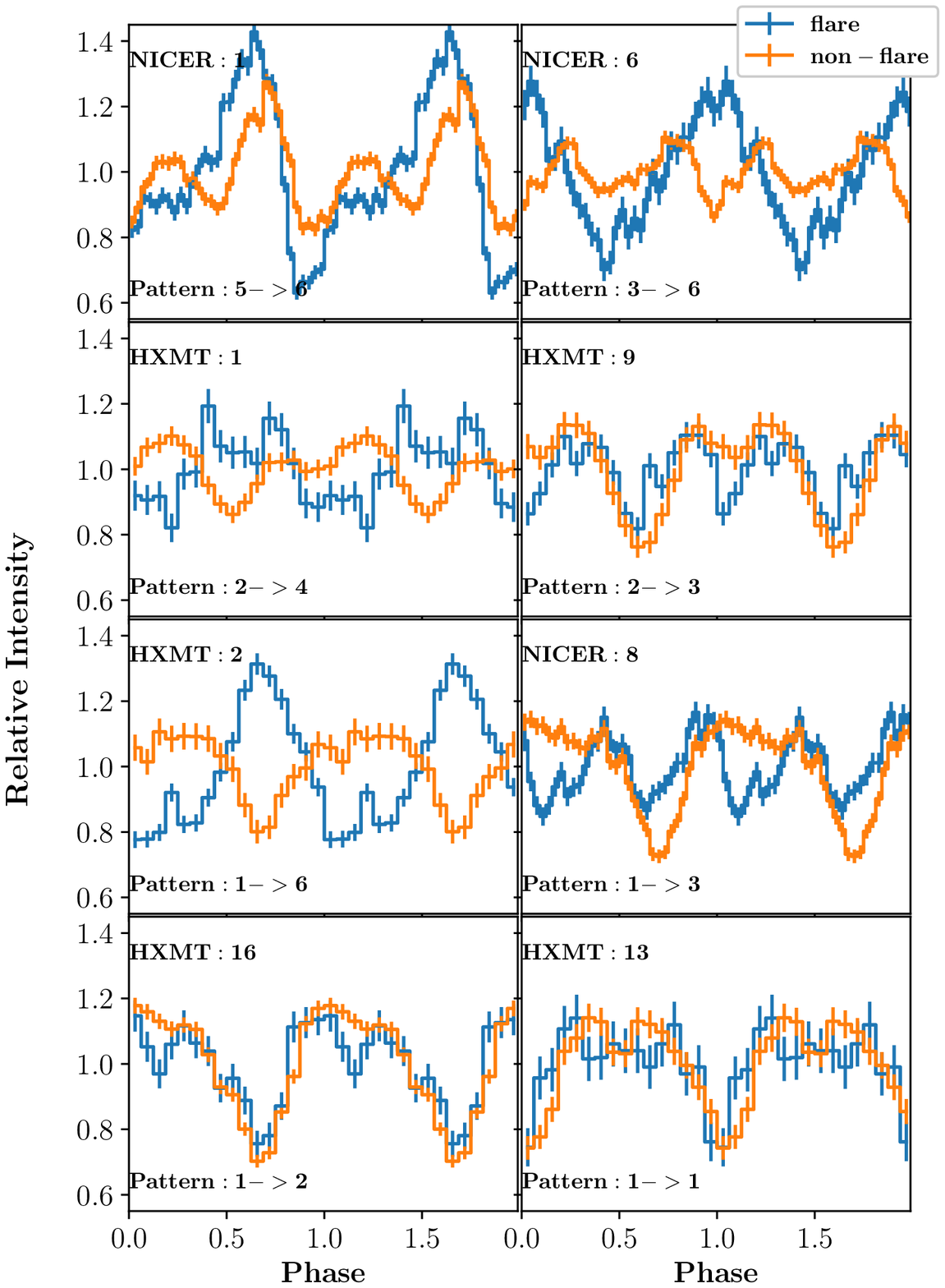}
	\includegraphics[width=0.425\columnwidth]{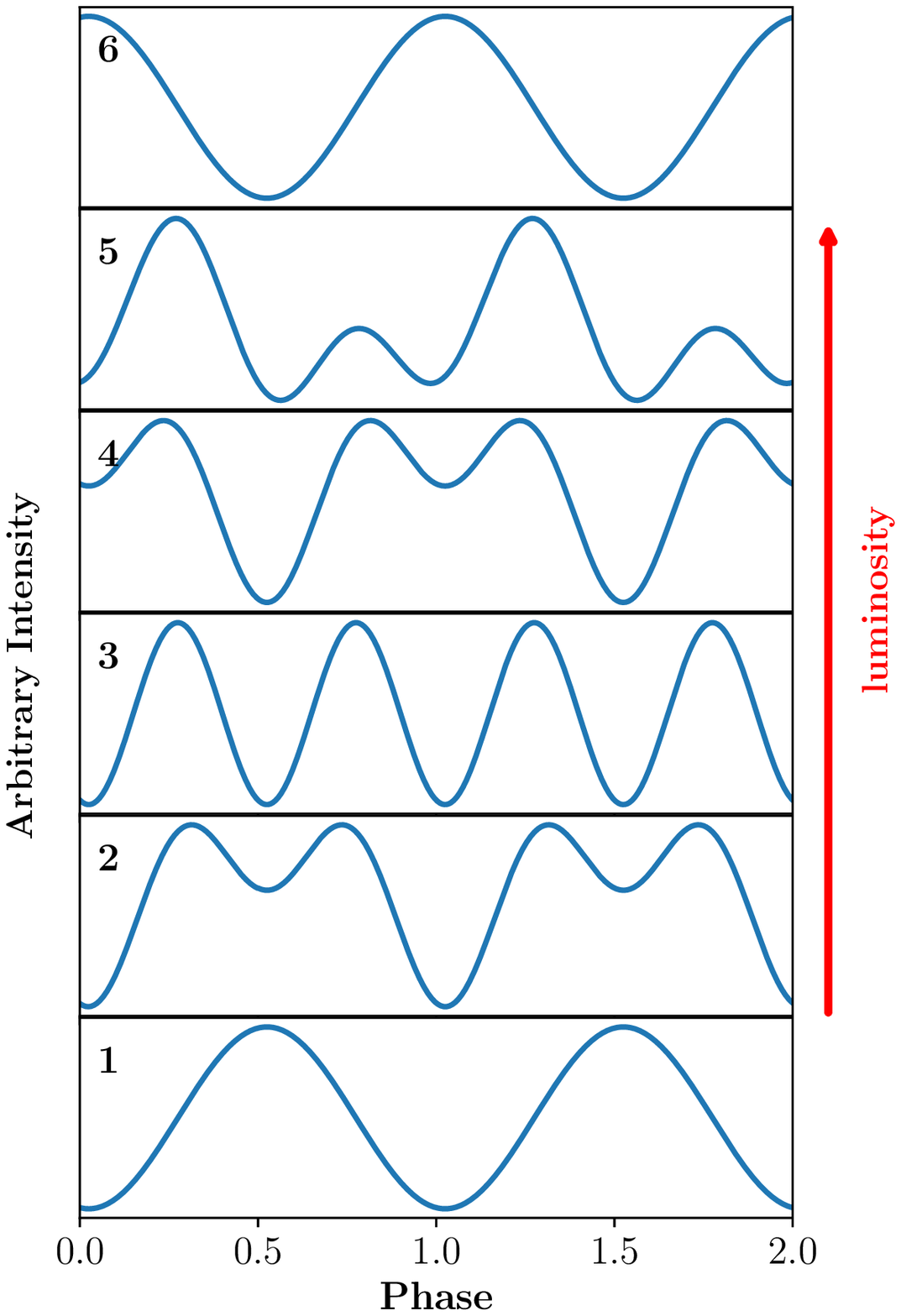}
    \caption{
    \textit{Left panels}: representative examples of pulse profiles of flares (blue) and non-flares (orange) observed with \textit{NICER} at 1-10\,keV and \textit{Insight-HXMT}/ME at 10-20\,keV (more information shown in Tables~\ref{tab:nicer} and~\ref{tab:hxmt}).
    The number at the top-left corner is the No. in our flare sample.
    \textit{Right panels}: a schematic of the evolution of pulse profiles with luminosity.
    The pulse phase corresponds to the geometrical angle of the rotation of the neutron star, and changes of pulse profiles represent the evolution of the beaming patterns.
    }
    \label{fig:sketch}
\end{figure*}

We studied the pulse profile shape during the flares in comparison to that of the nearby non-flare emissions considered within $\sim$ 300\,s (100 cycles) close to the flares.
We found that the former was generally quite different from the latter although in some cases they are similar.
We note that the shapes of pulse profiles during flares are quite stable (e.g., see Figure~\ref{fig:flares_sub}). Therefore we studied them by combining all available events during flares to increase statistics.
Here considering the complexity of pulse profiles, we report all results (including the observational time, count rates, the phase shift, patterns and other information) in Appendix and only show a selected sample in the main text for clarity.
This sample consists of representative examples, which contains core information in this paper and is shown in Figure~\ref{fig:sketch}.

Around the outburst peak, the pulse profile during flares shows only one broad peak (e.g., NICER No. 1 and 6 flares shown in Figure~\ref{fig:sketch}), while the non-flare pulse profiles exhibit a clear evolution from one main peak together with another small peak to two peaks as mentioned already (i.e., Figure~\ref{fig:ppev}).
At low outburst luminosities, the pulse profile of non-flares shows one broad peak with a notch in some cases, while the flaring pulse profiles are complex.
The shape of flares' pulse profiles could be a strong peak (e.g., \textit{HXMT} No. 1 and 2), or two separated peaks (e.g., \textit{NICER} No. 8 and \textit{HXMT} No. 9), or similar to that of the corresponding non-flare (e.g., \textit{HXMT} No. 13 and 16) (Figure~\ref{fig:sketch}).
We note that sometimes the pulse profiles during flares resemble those of the non-flares shown in a brighter outburst state, for example, the flare No. 8 and the non-flare No. 6 observed with \textit{NICER} (Figure~\ref{fig:sketch}).

We note that the reference time for Phase=0 is the same when we produced pulse profiles for flares and the nearby non-flares.
Therefore, the phase change of the peak of pulse profiles directly reflects the change of the view angle emitting most emissions with respect to the polar cap.
We studied the averaged pulse shift \footnote{When pulse profiles have two symmetric peaks (i.e., pattern 3), the "pulse shift" can not be clearly defined. But for the consistency we also include them in the following analysis. We note that our results will not be influenced if excluding them.} of pulse profiles between flares and the corresponding non-flares (i.e., the phase change of the peak of pulse profiles), which directly indicates the changes of beaming patterns on the surface of the neutron star.
Thus we calculated the pulse shift using cross-correlation technique and the error via Monte-Carlo simulations (Figure~\ref{fig:shift}).
Here a positive value means that the flare phase precedes the non-flare phase.
We found that such a shift happens only in intermediate luminosities of the outburst, and does not appear in the bright and faint states.
The pulse shifts are generally between 0.40-0.75, most of them are $\sim$ 0.5 apart from No. 5-7 observed with \textit{NICER}.
Considering that the pulse phase indicates a rotational angle of the neutron star, a phase shift suggests significant changes of the beaming pattern around the polar cap.

\begin{figure}
\centering
	\includegraphics[width=0.6\columnwidth]{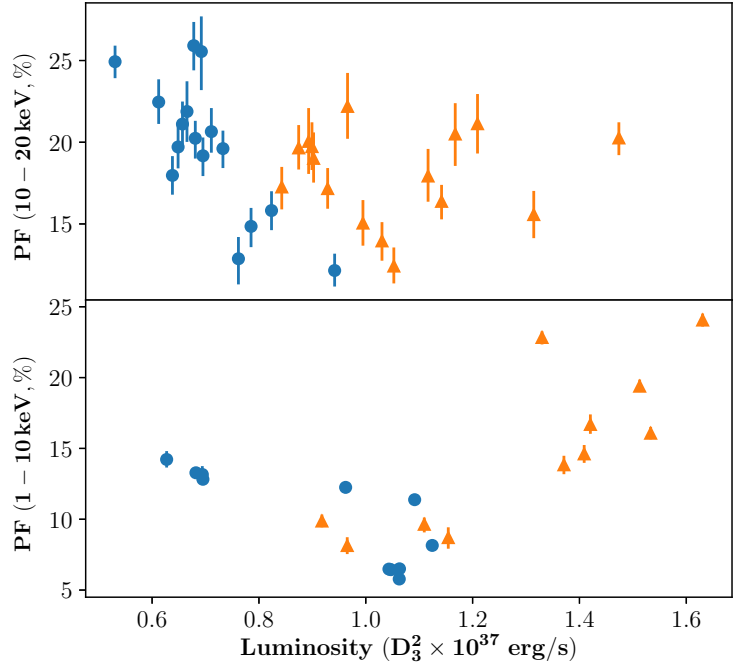}
    \caption{
    The rms pulsed fraction of flares (orange) and non-flares (blue) observed with \textit{Insight-HXMT} at 10-20\,keV and \textit{NICER} at 1-10\,keV.
    }
    \label{fig:PF}
\end{figure}

In addition, we studied the evolution of the root mean squared pulsed fraction \citep[Equation. 1 in][]{Wilson2018} of flares and non-flares in the decay phase of the outburst using \textit{NICER} (1-10\,keV) and \textit{Insight-HXMT}/ME (10-20\,keV) observations (Figure ~\ref{fig:PF}).
The pulsed fraction decreases with the increasing of luminosity ($L$) until $L$ $\sim$ $D_{3}^2$ $\times \rm 10^{37}\, erg/s$, where $D_{3}$ is the distance in the units of 3\,kpc, and remains constant or increases at higher luminosities.
Here $L$ is estimated by fitting the \textit{Insight-HXMT} spectrum with a phenomenological cutoff powerlaw model in the energy band of 1-20\,keV, or deduced from the \textit{NICER} count rate by multiplying a bolometric factor obtained from simultaneous observations with \textit{Insight}-HXMT.

\section{discussion}
We studied the hardness and the timing properties of 4U 1901+03 during the recent outburst in 2019 using observations obtained by \textit{NICER} and \textit{Insight-HXMT}.
We found that the hardness remained consistent at the outburst peak until $\sim$ MJD\,58550 and gradually decreased afterwards.
Around the same time, the pulse profile shape shows dramatic changes from a pattern featuring a main peak together with another small peak in the bright state to a simple broad peak in the faint state.
In the transition state the pulse profile is complex and highly variable.
The observed evolution of the hardness and the pulse profile is quite similar to what was reported for the previous outburst in 2003 \citep{Chen2008,Lei2009,Reig2016}.

X-ray accreting pulsars have different accretion regimes associated with luminosity.
If the luminosity is larger than a critical value $L_{\rm crit}$ (the super-critical regime), the accreting matter is decelerated by the vertical radiation flux that is high enough to stop accretion.
In this case, an optically thick accretion column is formed above the polar cap.
The emission then escapes through the walls of the column, resulting in a "fan-like" beaming pattern of the radiation.
On the other hand, if the luminosity is smaller than $L_{\rm crit}$ (the sub-critical regime), the radiation flux cannot completely stop the accreting matter and the rest of the kinetic energy of the particles is released in the neutron star atmosphere. In this case, a bright spot on the neutron star surface forms and the radiation escapes along the magnetic line, forming a "pencil beam" radiation.
We note that the beam pattern could be much more complex than a simple pencil/fan beam \citep[see, e.g.,][]{Kraus1995,Kraus1996,Sasaki2012,Becker2012,Mushtukov2018}.
Generally, the observational evidence of accretion regimes comes from changes of pulse profiles, spectral shapes, and cyclotron line energies \citep[e.g.,][]{Parmar1989, Klochkov2011, Reig2013, Becker2012, Mushtukov2015, Doroshenko2017}.
Thus, in 4U 1901+03 the complex variation of the pulse profile was naturally interpreted as a luminosity-dependent emission beam of the pulsar, a mixture of the "fan" and "pencil" beams: being dominated by the former (latter) at a high (low) luminosity \citep{Chen2008}.
The change of the hardness might also suggest a transition between the super-critical and sub-critical states although no distinct branches in the hardness-intensity diagram were found \citep{Reig2016}.

We note that such changes can occur on short time scales, i.e., during the flares observed in the decaying phase of the outburst in 4U 1901+03.
We found that pulse profiles of flares are generally distinct from those of corresponding non-flares, although sometimes they do show a similar shape.
Instead, they are quite similar to non-flares' pulse profiles exhibited during a brighter outburst state.
We note that flares are not uncommon in X-ray pulsars, such as A0535+26, GRO J1744-28, SMC X-1, Swift J1626.6-5156, LMC X-4 and A0538-66 \citep{Caballero2008, Woods2000, Moon2003, Reig2008, Shtykovsky2018, Ducci2019}.
However, flares in these sources are generally much brighter than those in 4U 1901+03, and lack a direct link to non-flare emissions.
A clear phase shift between the flare and the nearby non-flare emissions was only reported in GRO J1744-28 which, however, has two main differences from 4U 1901+03: first, the phase shift in GRO J1744-28 is much smaller; second, the phase shift in GRO J1744-28 did not recover completely after flares and thus likely has other origins \citep{Stark1996}.
We emphasise that the pulse profile shape of flares in 4U 1901+03 is therefore quite unique, and it is the first time that this type of variability of pulse profiles is reported.

By comparing the pulse profiles of the source during the flares with the non-flare ones, we propose a unified picture of the evolution of beaming patterns with luminosity ($L$) in 4U 1901+03 (see Figure~\ref{fig:sketch}, right panel):\\
1) when the luminosity is low ($\lesssim  0.7 \times 10^{37} D_{3}^2 \,{\rm ergs\ s^{-1}}$), the pulse profile has only one peak ;\\
2) a notch appears and gradually becomes more significant at $L \sim 0.8 \times 10^{37} D_{3}^2 \,{\rm ergs\ s^{-1}}$. \\
3) a pulse profile becomes two symmetric peaks at $L \sim 1.0 \times 10^{37} D_{3}^2 \,{\rm ergs\ s^{-1}}$;\\
4) a pulse profile similar to pattern 2 appears again at a higher luminosity but with a shifted phase (see the flare No 1. of \textit{Insight-HXMT}) ($L \sim 1.2 \times 10^{37} D_{3}^2 \,{\rm ergs\ s^{-1}}$);\\
5) around the outburst peak, the pulse profile becomes asymmetric, showing a main peak plus another small peak ($L \sim 1.3 \times 10^{37} D_{3}^2 \,{\rm ergs\ s^{-1}}$);\\
6) for flares at the outburst peak, which have the highest luminosity, a simple one-peak profile appears again, which is significantly shifted compared to the case of the lowest luminosity ($L \gtrsim 1.4 \times 10^{37} D_{3}^2 \,{\rm ergs\ s^{-1}}$).

{The pulse profile evolution of non-flares with luminosity has been shown in Figure~\ref{fig:ppev}, in which however patterns between 2 and 4 can not be distinguished,
because long gaps between observations and the evolution of pulse profiles prevent us from obtaining absolute pulse phases.
In other words, although pulse profiles before and after $\sim$ MJD 58555 seem to be similar, they do have different beam patterns.
A transition from pattern 2 to 4 can only be investigated by comparing the flare and the corresponding non-flare pulse profiles when pulse phases can be directly compared.
}

\begin{figure}
\centering
	\includegraphics[width=0.6\columnwidth]{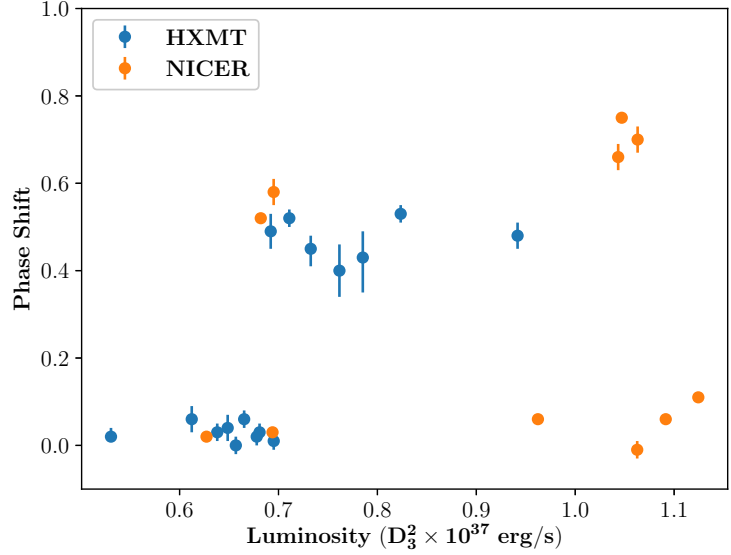}
    \caption{
    Non-flare luminosity  vs. phase shift during flares.
    }
    \label{fig:shift}
\end{figure}

We note that, during the flares, pulse profiles could be changed from one pattern to another.
A pulse shift shown in Figure~\ref{fig:shift} during the flares is observed if a flare has a luminosity with a pattern 4-6 while the corresponding non-flare stays in the range of a pattern 1-2, e.g., \textit{HXMT} flares No. 1 and 2 in  Figure~\ref{fig:sketch}.
Such a shift is a direct evidence of the transition in beam patters of the pulsar, expected for the transition between the sub-critical and super-critical regimes, i.e., where the radiation is parallel and perpendicular to magnetic field lines, respectively \citep{Basko1976,Becker2012, Mushtukov2015}.
On the other hand, the shift is not significant if the pattern changes within the same accretion regime, e.g., \textit{NICER} No. 1 and \textit{Insight-HXMT} No. 13 and 16 in Figure~\ref{fig:sketch}.

We note that the most important state is the intermediate state (pattern 3; two symmetric peaks), which corresponds to the transition of the outburst as mentioned above, i.e., between two dashed lines in Figure~\ref{fig:lc}.
This pattern appears in both flares and non-flares (Figure~\ref{fig:sketch}).
The inferred luminosity range is shown as a green strip
\footnote{Here we approximately calculated that 1 \textit{NICER} (\textit{Insight-HXMT}/ME) cts/s $\sim$ 0.5 (2.9)\,mCrab at 15-50\,keV.}
 in Figure~\ref{fig:lc}, which perfectly coincides with the luminosity where the outburst shows the transition.
This indicates that the short-term and long-term flux variabilities could result in a similar accretion process on the polar cap.
As expected, the analysis of the short-term variability is an independent method to investigate the accretion physics and the radiation process on the polar cap in X-ray pulsars.

The critical luminosity ($L_{\rm c}$) between two accretion regimes, i.e., when the pattern 3 appears, is $\sim$  $10^{37}$ $\rm ergs\ s^{-1}$.
Here the luminosity is estimated by fitting broad band \textit{Insight-HXMT} spectra with a phenomenological model "cutoffpl" in the energy range of 1-20\,keV, assuming a distance of 3\,kpc measured by \textit{Gaia} \citep{Bailer2018}.
\citet{Becker2012} suggests that the critical luminosity is associated with the magnetic field as $L_{\rm c} \sim 1.5 \times 10^{37}B_{12}$\,$\rm ergs\ s^{-1}$, where $B_{12}$ is the magnetic field strength in units of $10^{12}$\,G.
Thus the inferred magnetic field of the source is $\sim$ $7.5\times10^{11}$\,G.
In high magnetised neutron stars, CRSFs are usually found at $E_{\rm cyc} \approx 11.6\,B_{12}$\,keV \citep[for details, see, e.g.,][]{Staubert2019}.
Therefore, a cyclotron line with a centroid energy of around 9\,keV is predicted.
{This deduction coincides with the previous suggestion that the prominent "10\,keV feature" as modeled by a absorption-like line is a CRSF \citep{Reig2016,Mereminskiy2019}.
}
But we caution that it is also possible that this feature arises from limitations of simple phenomenological models used to describe the continuum \citep[see, e.g.,][]{Coburn2002}.
On the other hand, \citet{Mushtukov2015} proposed a different relation between the critical luminosity and the magnetic field, which suggests that the CRSF energy is $\sim$ 30\,keV.
This deduction is consistent with the 30\,keV absorption feature reported by \citet{Coley2019} at a confidence level of 99.99\%.
The question of the CRSF in 4U 1901+03 is still open.
Further observations are required to confirm this 30\,keV absorption feature and search for potential CRSF harmonics.


%--------------------------------------------------------------------------------------------

%---------------------------------------------------------------------------------------------

\section*{Acknowledgements}
This work made use of the data from the Insight-HXMT mission, a project funded by China National Space Administration (CNSA) and the Chinese Academy of Sciences (CAS). The Insight-HXMT team gratefully acknowledges the support from the National Program on Key Research and Development Project (Grant No. 2016YFA0400800) from the Minister of Science and Technology of China (MOST) and the Strategic Priority Research Program of the Chinese Academy of Sciences (Grant No. XDB23040400). The authors thank supports from the National Natural Science Foundation of China under Grants No. 11503027, 11673023, 11733009,
U1838201, U1838202 and U1938103.
We acknowledge the use of public data and products from the \textit{Swift} and \textit{NICER} data archive.
JL thanks the German Academic Exchange Service (DAAD, project57405000) for travel grants. VS, ST and VD acknowledge the support from the Russian Science Foundation grant 19-12-00423. VS thanks the Deutsche Forschungsgemeinschaft (DFG) grant WE 1312/51-1.

%%%%%%%%%%%%%%%%%%%%%%%%%%%%%%%%%%%%%%%%%%%%%%%%%%


\bibliographystyle{mnras}
\bibliography{mybibtex}
\clearpage


% \begin{thebibliography}{99}
% \bibitem[\protect\citeauthoryear{Author}{2012}]{Author2012}
% Author A.~N., 2013, Journal of Improbable Astronomy, 1, 1
% \bibitem[\protect\citeauthoryear{Others}{2013}]{Others2013}
% Others S., 2012, Journal of Interesting Stuff, 17, 198
% \end{thebibliography}

%%%%%%%%%%%%%%%%%%%%%%%%%%%%%%%%%%%%%%%%%%%%%%%%%%

%%%%%%%%%%%%%%%%% APPENDICES %%%%%%%%%%%%%%%%%%%%%

%%%%%%%%%%%%%%%%%%%%%%%%%%%%%%%%%%%%%%%%%%%%%%%%%%

\appendix
\section{} \label{appendix}

\begin{figure}
\centering
	\includegraphics[width=0.75\columnwidth]{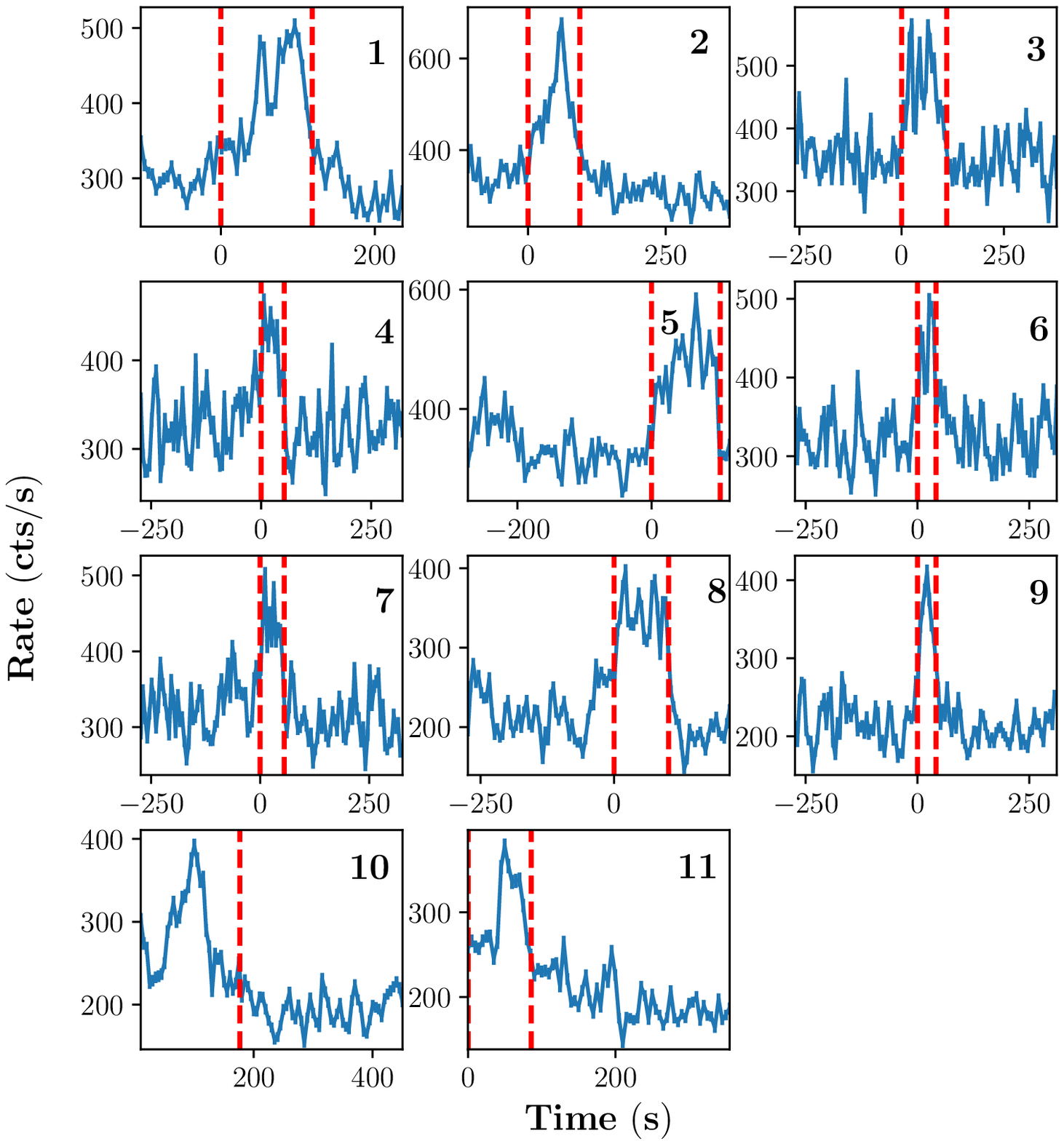}
    \caption{Flares observed with \textit{NICER} at 1-10\,keV. The red vertical lines represent the starting and ending points of flares.}
    \label{fig:nicer}
\end{figure}

\begin{figure}
\centering
	\includegraphics[width=0.75\columnwidth]{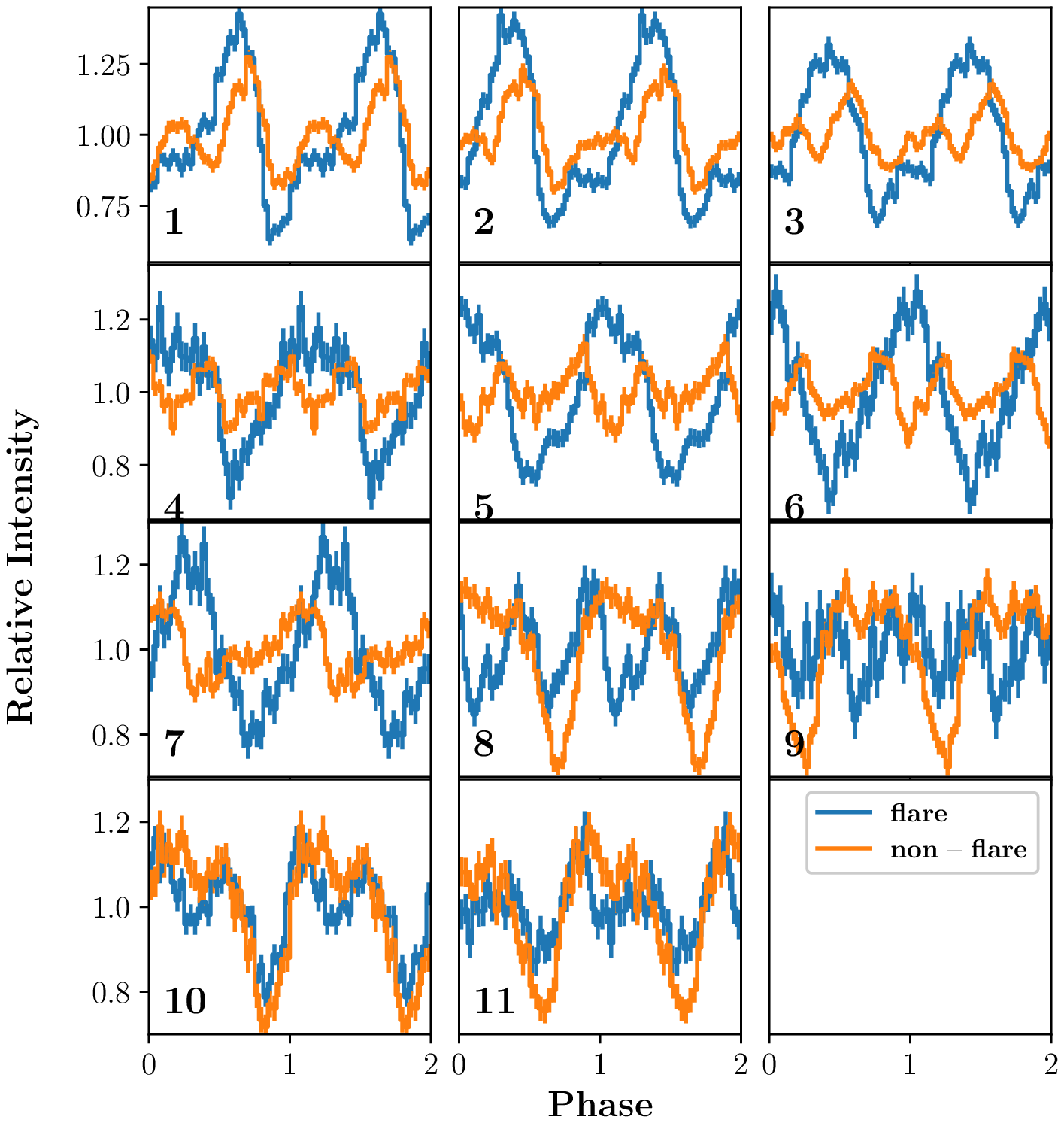}
    \caption{Pulse profiles of the flare (blue) and non-flare (orange) emissions in the energy range of 1-10\,keV observed with \textit{NICER}.}
    \label{fig:nicerPP}
\end{figure}

\begin{figure}
\centering
	\includegraphics[width=0.75\columnwidth]{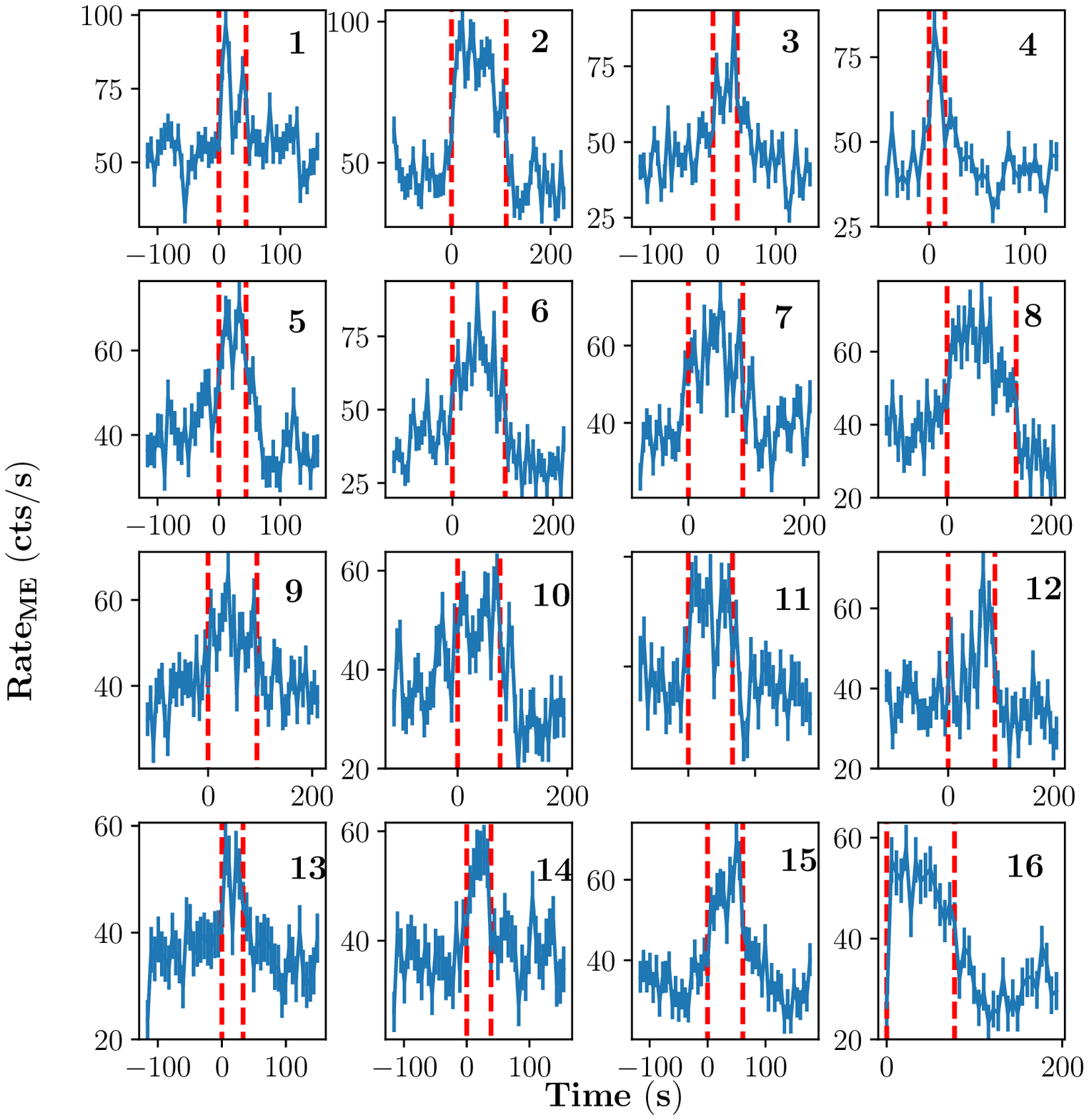}
    \caption{Flares observed with \textit{Insight-HXMT}/ME at 10-20\,keV. The red vertical lines represent the starting and ending points of flares. }
    \label{fig:hxmt}
\end{figure}

\begin{figure}
\centering
	\includegraphics[width=0.75\columnwidth]{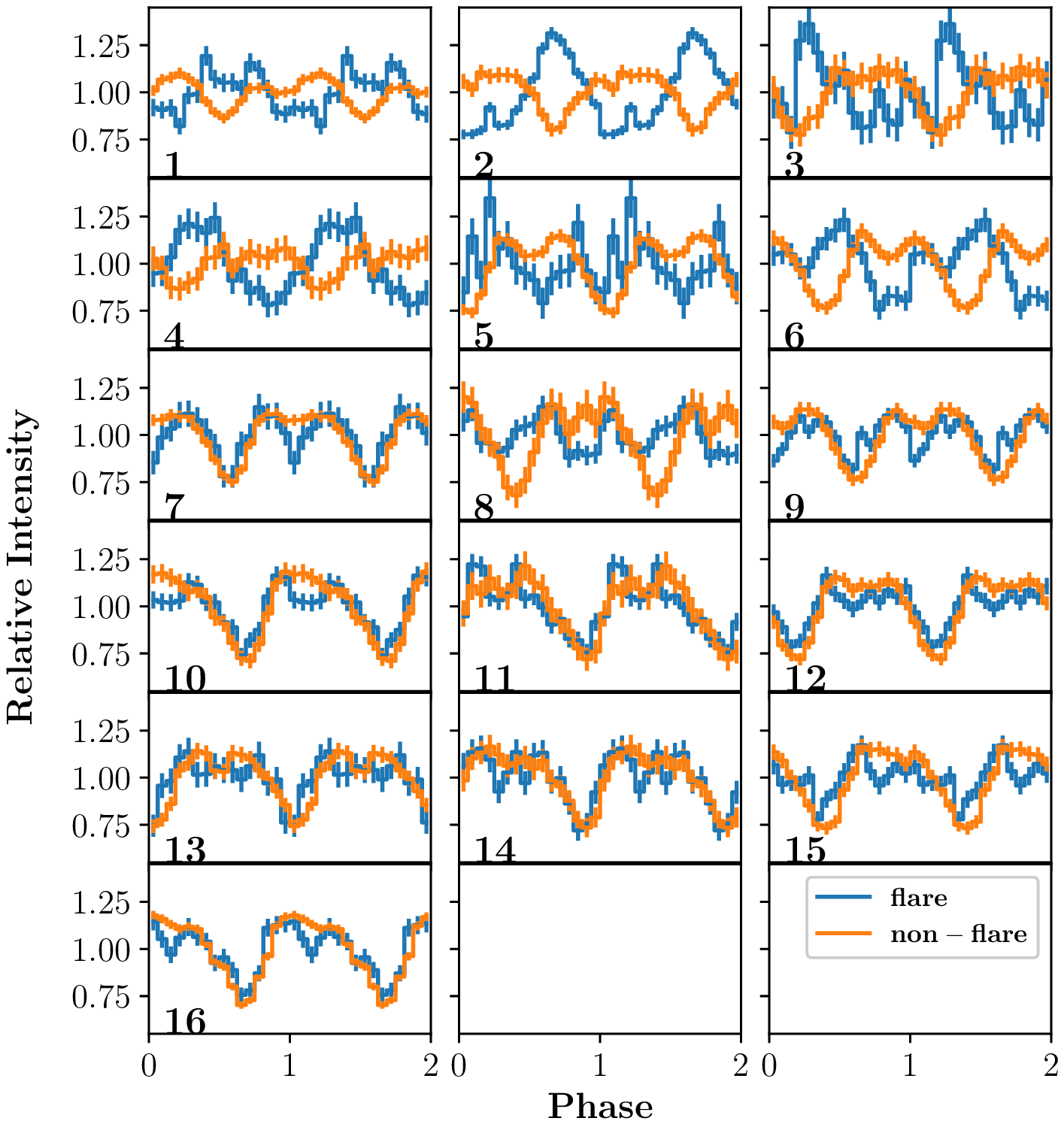}
    \caption{Pulse profiles of flares (blue) and non-flares (orange) emissions in the energy range of 10-20\,keV observed with \textit{Insight-HXMT}.}
    \label{fig:hxmtPP}
\end{figure}

\begin{table*}
\centering
\caption{The columns denote the No., the observational ID of \textit{NICER}, the time when flares occurred, the count rate of flare and non-flare emissions at 1-10\,keV, the duration of flares, the phase shift between flare and non-flare emissions, and beaming patterns. "?" represents the pattern that can not be identified unambiguously.}
	\begin{tabular}{cccccccc}
		\hline
		No. & ObsId & Time & $\rm Rate_{\rm NF}$ & $\rm Rate_{\rm F}$ & $T_{\rm F}$ &  Phase Shift  & Pattern \\
		& & (MJD) & (cts/s) & (cts/s) & (s) &     & \\ % & (\%)  & (\%)  \\
		\hline
1  & 1200570104    & 58535.44 & $294.52\pm1.13$  & $407.04\pm1.85$  & 118.73 & $0.06_{-0.01}^{+0.01}$  & 5->6\\ % 
2  & 1200570112    & 58543.55 & $334.07\pm1.15$  & $499.07\pm2.31$  & 93.93  & $0.06_{-0.01}^{+0.01}$  & 5->6\\ % 
3  & 2200570104    & 58549.60 & $344.15\pm1.12$  & $462.99\pm2.05$  & 110.51 & $0.11_{-0.01}^{+0.01}$  & 5->6\\ % 
4  & 2200570110    & 58556.29 & $325.24\pm1.09$  & $419.59\pm2.83$  & 52.46  & $-0.01_{-0.02}^{+0.02}$ & 4->6\\ % 
5  & 2200570111    & 58557.64 & $319.35\pm1.43$  & $469.29\pm2.14$  & 102.16 & $0.66_{-0.03}^{+0.03}$  & 3 ->6\\ % 
6  & 2200570112    & 58558.93 & $325.37\pm1.09$  & $434.75\pm3.24$  & 41.42  & $0.70_{-0.03}^{+0.03}$  & 3->6\\ % 
7  & 2200570114    & 58561.51 & $320.46\pm1.08$  & $431.20\pm2.79$  & 55.23  & $0.75_{-0.01}^{+0.01}$  & ?->6\\ % 
8  & 2200570118    & 58585.55 & $208.76\pm0.90$  & $339.58\pm1.82$  & 102.19 & $0.52_{-0.01}^{+0.01}$  & 1->3\\ % 
9  & 2200570118    & 58585.24 & $212.75\pm0.88$  & $353.33\pm2.92$  & 41.43  & $0.58_{-0.03}^{+0.03}$  & 1->3\\ % 
10 & 2200570118    & 58585.30 & $191.97\pm1.18$  & $280.93\pm1.29$  & 176.75 & $0.02_{-0.01}^{+0.01}$  & 1->2\\ % 
11 & 2200570120    & 58589.66 & $212.40\pm1.23$  & $295.44\pm1.86$  & 85.64  & $0.03_{-0.01}^{+0.01}$  & 1->2\\ % 
		\hline
	\end{tabular}
\label{tab:nicer}
\end{table*}

\begin{table*}
\centering
\caption{The columns denote the No., the observational ID of \textit{Insight-HXMT}, the time when flares occurred, the count rate of flare and non-flare emissions at 10-20\,keV, the duration of flares, the phase shift between flare and non-flare emissions, and beaming patterns. "?" represents the pattern that can not be identified unambiguously.}
\begin{tabular}{cccccccc}
\hline
No. & ObsId & Time & $\rm Rate_{\rm NF}$ &   $\rm Rate_{\rm F}$ & $T_{\rm F}$ &  Phase Shift   & Pattern  \\ %
& & (MJD) & (cts/s) &  (cts/s) & (s)  &    & \\%& (\%)  &  (\%)  \\
\hline
1  & P021100600102 & 58563.57 & $52.60 \pm 1.05$ & $73.42 \pm 1.22$ & 44.19  & $0.48_{-0.03}^{+0.03}$  & 2->4\\ % 
2  & P021100600104 & 58563.86 & $46.00 \pm 1.01$ & $82.32 \pm 0.86$ & 110.48 & $0.53_{-0.02}^{+0.02}$  & 1->6\\ %
3  & P021100600106 & 58564.06 & $43.86 \pm 0.97$ & $67.54 \pm 1.24$ & 38.67  & $0.43_{-0.08}^{+0.06}$  & 1->?\\ % 
4  & P021100600106 & 58564.13 & $42.54 \pm 2.14$ & $65.21 \pm 1.72$ & 16.57  & $0.40_{-0.06}^{+0.06}$  & 1->6\\ % 
5  & P021100600501 & 58577.77 & $40.92 \pm 0.95$ & $62.37 \pm 1.12$ & 44.19  & $0.45_{-0.04}^{+0.03}$  & 2->?\\ %
6  & P021100600302 & 58574.01 & $39.71 \pm 0.96$ & $63.78 \pm 0.76$ & 104.95 & $0.52_{-0.02}^{+0.02}$  & 2->4\\ %
7  & P021100600601 & 58579.69 & $38.83 \pm 0.95$ & $57.55 \pm 0.76$ & 93.90  & $0.01_{-0.02}^{+0.02}$  & 1->3\\ %
8  & P021100600302 & 58573.94 & $38.66 \pm 0.98$ & $58.79 \pm 0.65$ & 132.57 & $0.49_{-0.04}^{+0.04}$  & 1->3\\ %
9  & P021100600301 & 58573.79 & $38.03 \pm 0.95$ & $51.87 \pm 0.72$ & 93.91  & $0.03_{-0.02}^{+0.02}$  & 2->3\\ % 
10 & P021100601301 & 58596.45 & $37.87 \pm 0.94$ & $50.23 \pm 0.78$ & 77.33  & $0.02_{-0.02}^{+0.02}$  & 1->2\\ % 
11 & P021100601101 & 58589.10 & $37.16 \pm 1.28$ & $50.40 \pm 0.87$ & 66.28  & $0.06_{-0.02}^{+0.02}$  & 1->2\\ %
12 & P021100600901 & 58586.78 & $36.69 \pm 0.94$ & $47.07 \pm 0.73$ & 88.38  & $0.00_{-0.02}^{+0.02}$  & 1->1\\ % 
13 & P021100600801 & 58583.28 & $36.23 \pm 1.00$ & $49.88 \pm 1.23$ & 33.14  & $0.04_{-0.03}^{+0.03}$  & 1->1\\ %
14 & P021100601101 & 58589.10 & $35.64 \pm 0.94$ & $53.95 \pm 1.28$ & 38.67  & $0.03_{-0.02}^{+0.02}$  & 1->2\\ % 
15 & P021100601201 & 58594.52 & $34.20 \pm 0.92$ & $55.56 \pm 0.92$ & 60.76  & $0.06_{-0.03}^{+0.03}$  & 1->2\\ % 
16 & P021100601301 & 58596.51 & $29.64 \pm 0.98$ & $48.84 \pm 0.77$ & 77.33  & $0.02_{-0.01}^{+0.02}$  & 1->2\\ %
\hline
\end{tabular}
\label{tab:hxmt}
\end{table*}

% Don't change these lines
\bsp	% typesetting comment
\label{lastpage}
\end{document}